\documentclass[prd,a4paper,nofootinbib,
showpacs, 
]{revtex4}
\usepackage{graphicx}
\usepackage{amsfonts}
\usepackage{amssymb}
\def\eq#1{{eq.~(\ref{#1})}}

\def\vev#1{\left\langle #1\right\rangle}

\def\hbar{\hspace{0pt}\raisebox{1pt}{$-$} \hspace{-7pt} h}

\def\5{\overline 5}

\newcommand{\be}{\begin{equation}}
\newcommand{\ee}{\end{equation}}
\newcommand{\bea}{\begin{eqnarray}}
\newcommand{\eea}{\end{eqnarray}}
\newcommand{\nn}{\nonumber}

\begin{document}
\title[]{Quark contact interactions at the LHC
}
\date{\today}
\author{F. Bazzocchi$^{\dag\ddag}$}
\author{U. De Sanctis$^{\dag *}$}
\author{M. Fabbrichesi$^{\ddag}$}
\author{A. Tonero$^{\dag\ddag}$}
\affiliation{$^{\ddag}$INFN, Sezione di Trieste}
\affiliation{$^{*}$INFN, Gruppo collegato di Udine} 
\affiliation{$^{\dag}$SISSA, via Bonomea 265, 34136 Trieste, Italy}

\begin{abstract}
\noindent   Quark contact interactions are an important signal of new physics.  We introduce a  model in which the presence of a symmetry  protects these new interactions from giving large corrections in flavor changing processes at low energies. This minimal model provides the basic set of operators which must be considered to contribute to the high-energy processes. To discuss their experimental signature in  jet pairs produced in proton-proton colllisions, we simplify the number of possible operators down to two. We show (for a representative integrated luminosity of 200 pb$^{-1}$ at $\sqrt{s}= 7$ TeV) how the presence of two operators  significantly modifies  the bound on the characteristic energy scale of the contact interactions which is obtained by keeping a single operator.  

\end{abstract}

\maketitle
\vskip1.5em
\section{Motivations} 
\label{sec:mot}

Fermions like quarks can be made to interact directly---that is, without the exchange of an intermediate particle---by simply adding to the standard model (SM) lagrangian four-quark contact terms like, for example,
\be
\frac{2 \pi }{\Lambda^2} \;\bar \psi_{L}\gamma^\mu\psi_{L}\bar \psi_{L}\gamma_\mu\psi_{L}\, , \label{fermi}
\ee 
where  $2 \pi $ gives the strength, $\Lambda$ the characteristic energy scale of this new interaction and the quark fields $\psi$ are taken to be left-handed. 

Because of the non-renormalizability of the term in \eq{fermi}, such an operator is often thought as the low-energy effective approximation of a renormalizable lagrangian in which heavy particles are exchanged. These heavy particles can be of many different kinds, each kind giving rise to an effective operator with different color, flavor and Dirac structure. 
Sometimes these heavy states are thought of as a substructure of the quarks themselves and in this case the contact interactions are presented as evidence for quark compositeness. More in general, the heavy states represent new physics which lives at an energy scale that is  too high to manifest itself with the production of the new states either as intermediate resonances or in chain-decay processes, and the effect of which can only be seen  by the effective operators of the contact interactions~\cite{Eichten}.
Non-renormalizability is not a problem in models in which the couplings run toward a ultraviolet  fixed point. In these asymptotically safe (AS) models~\cite{AS} the contact interactions in \eq{fermi} can be considered as fundamental. It has recently been shown that indeed such operators arise in a natural manner in AS models of the weak interactions and a search for their presence could provide an important experimental clue~\cite{noi}.

In any case, be the contact interactions fundamental or  remnant of new physics at higher energies, the search for their existence and the bounds on their characteristic energy scale is important. The LHC has already provided us with new 
constraints~\cite{atlas,cms}. 
These constraints were derived  by assuming the existence of only one kind of contact term, namely that in \eq{fermi} in which the operator is given by the product of two left-handed quark currents.
This has become the standard practice following \cite{Eichten} because it is simple. Unfortunately, this is too restrictive an assumption and the significance of an analysis based on it  is unavoidably weakened.  

It is easy to imagine a great variety of different operators contributing to the quark contact interactions. The problem is that this variety is  constrained by  stringent bounds on flavor physics at low energies~\cite{martinelli}. We do not want to track down every and each operator for its possible low-energy effect and, in order to provide a minimal model, we impose a symmetry on the possible contact interactions which makes them safe with respect to these low-energy constraints. This model defines a basic set of operators whose size is not severely constrained by flavor physics and the existence of which can be tested in high-energy processes. It is also general enough to make the bound on the characteristic energy scale $\Lambda$ realistic.

In principle, all these operators should be entered in the analysis, each with a different strength. To make the analysis manageable, we simplify further down to two the number of possible operators. This skeleton model is sufficient in showing---in  jet pairs produced in proton-proton colllisions at the LHC for a representative integrated luminosity of 1 fb$^{-1}$ at $\sqrt{s}= 7$ TeV---how the bound on the energy scale of the contact interactions is sensitive to the relative strengths of different  terms. This exercise shows   that at least a subset of  operators should be taken into account, current analyses based on a single operator cannot be considered as final, and  the actual bound on the characteristic energy scale $\Lambda$ is  weaker than reported~\cite{atlas,cms}.

\section{The minimal model}
\label{sec:model}

Let us for a moment consider the case of a single fermion family with quarks with the same mass. The most general four-quark interaction will depend on the overall symmetry we want to impose on the system. It could be $U(1)_Q$  or $SU(2)\times U(1)$ and in these cases we would have, respectively, 20  or 10 possible terms. Following the general idea that stronger  interactions are more symmetric than weaker ones, we want to be more restrictive and impose a larger group  $SU(2)_L\times SU(2)_R$  and a parity symmetry.

The complete set of $SU(2)_L\times SU(2)_R$ and parity invariant four-fermion operators is given by four independent terms:
\bea\label{4f}
{\cal L}_{\psi^4} &=& \lambda_{1} \, \left( \bar \psi_L^{ia} \psi_R^{ja} \bar \psi_R^{jb} \psi_L^{ib}\right)+ \lambda_{2} \, \left(\bar \psi_L^{ia} \psi_R^{jb}  \bar \psi_R^{jb} \psi_L^{ia} \right)\nn\\
&+& \lambda_{3} \, \left( \bar \psi_L^{ia}\gamma_\mu \psi_L^{ia} \bar \psi_L^{jb}\gamma^\mu \psi_L^{jb} +\bar \psi_R^{ia}\gamma_\mu \psi_R^{ia} \bar \psi_R^{jb}\gamma^\mu \psi_R^{jb}\right)\nn\\
&+& \lambda_{4} \, \left( \bar \psi_L^{ia}\gamma_\mu \psi_L^{ib} \bar \psi_L^{jb}\gamma^\mu \psi_L^{ja} +\bar \psi_R^{ia}\gamma_\mu \psi_R^{ib} \bar \psi_R^{jb}\gamma^\mu \psi_R^{ja}\right) \, 
\eea
where the dimensional coefficients $\lambda_n$ can be written as  $2 \pi /\Lambda_n^2 $ in terms of the characteristic energies $\Lambda_n$. Indices $i,j$ and $a,b$ are $SU(2)$ and color indices, respectively. 

The operators in \eq{4f} must be considered together with  the Yukawa term of the standard model:
\be
\label{lagyuk}
- \frac{2h}{f} \big( \bar \psi_L^{ia} U^{ij} \psi_R^{ja} + \mathrm{h.c.}\big) \, .
\ee

For a realistic model we should consider the splitting between up and down-type quarks as well as the three SM families. This gives rise to a large proliferation of possible terms. At the same time, we must take care that the four-fermion operators do not  yield unwanted flavor-changing  neutral current (FCNC) processes with $\Delta F= 2$ which are strongly suppressed by the  experimental data.  

In such a realistic and most  general case the $\lambda_i$ in \eq{4f}  are 4 flavor-index tensors and the coefficient  $h$ in \eq{lagyuk} is a  matrix. Thus the four-fermion interaction lagrangian has $20$ 4-index tensor operators, that in general   are not simultaneously diagonalized with the Yukawa mass terms, thus giving rise to problematic operators such as
\be
(\bar{d}_L s_R)^2\,,
\ee
which affect for example meson oscillations~\cite{martinelli}. 

To prevent such operators and reduce the number of free parameters  one possibility is introducing a flavour symmetry. In the following we assume that left and right handed quarks transform as the fundamental representation of a continuous  family symmetry we choose to be $SU(3)_F$. In this case the set of four fermion operators  given in \eq{4f} becomes
\bea\label{4f2}
{\cal L}_{\psi^4}^\prime &=& \lambda_{1} \, \left( \bar \psi_{L_\alpha}^{ia} \psi_{R_\alpha}^{ja} \bar \psi_{R_\beta}^{jb} \psi_{L_\beta}^{ib}\right)+  \tilde{\lambda}_{1} \, \left( \bar \psi_{L_\alpha}^{ia} \psi_{R_\beta}^{ja} \bar \psi_{R_\beta}^{jb} \psi_{L_\alpha}^{ib}\right)\nn\\
&+& \lambda_{2} \, \left(\bar \psi_{L_\alpha}^{ia} \psi_{R_\alpha}^{jb}  \bar \psi_{R_\beta}^{jb} \psi_{L_\beta}^{ia} \right)+ \tilde{\lambda}_{2} \, \left(\bar \psi_{L_\alpha}^{ia} \psi_{R_\beta}^{jb}  \bar \psi_{R_\beta}^{jb} \psi_{L_\alpha}^{ia} \right)\nn\\
&+& \lambda_{3} \, \left( \bar \psi_{L_\alpha}^{ia}\gamma_\mu \psi_{L_\alpha}^{ia} \bar \psi_{L_\beta}^{jb}\gamma^\mu \psi_{L_\beta}^{jb} +\bar \psi_{R_\alpha}^{ia}\gamma_\mu \psi_{R_\alpha}^{ia} \bar \psi_{R_\beta}^{jb}\gamma^\mu \psi_{R_\beta}^{jb}\right)\nn\\
&+&\tilde{ \lambda}_{3} \, \left( \bar \psi_{L_\alpha}^{ia}\gamma_\mu \psi_{L_\beta}^{ia} \bar \psi_{L_\beta}^{jb}\gamma^\mu \psi_{L_\alpha}^{jb} +\bar \psi_{R_\alpha}^{ia}\gamma_\mu \psi_{R_\beta}^{ia} \bar \psi_{R_\beta}^{jb}\gamma^\mu \psi_{R_\alpha}^{jb}\right)\nn\\
&+& \lambda_{4} \, \left( \bar \psi_{L_\alpha}^{ia}\gamma_\mu \psi_{L_\alpha}^{ib} \bar \psi_{L_\beta}^{jb}\gamma^\mu \psi_{L_\beta}^{ja} +\bar \psi_{R_\alpha}^{ia}\gamma_\mu \psi_{R_\alpha}^{ib} \bar \psi_{R_\beta}^{jb}\gamma^\mu \psi_{R_\beta}^{ja}\right)\nn\\
&+& \tilde{ \lambda}_{4} \, \left( \bar \psi_{L_\alpha}^{ia}\gamma_\mu \psi_{L_\beta}^{ib} \bar \psi_{L_\beta}^{jb}\gamma^\mu \psi_{L_\alpha}^{ja} +\bar \psi_{R_\alpha}^{ia}\gamma_\mu \psi_{R_\beta}^{ib} \bar \psi_{R_\beta}^{jb}\gamma^\mu \psi_{R_\alpha}^{ja}\right)\,,
\eea
where $\alpha,\beta$ are $SU(3)_F$ flavor indices. The symmetry of the four-fermion interactions is thus $SU(2)_L\times SU(2)_R\times SU(3)_F$.

In this minimal model there are 8 arbitrary coefficients---the $\lambda_n$ and $\tilde \lambda_n$ in \eq{4f2}---each of them multiplying various operators which are different for flavor and color structure. In a numerical study, all coefficients should in principle be varied and the most relevant among the operators included.

For what concerns the Yukawa term given in \eq{lagyuk}, first of all we have to split up and down-type quarks: 
\be
\label{lag2}
- \frac{2h_{\alpha\beta}^u}{f} \big( \bar \psi_{L_\alpha}^{ia} U^{ij} P_u \psi_{R_\beta}^{ja} + \mathrm{h.c.}\big)   - \frac{2h_{\alpha\beta}^d}{f} \big( \bar \psi_{L_\alpha}^{ia} U^{ij} P_d \psi_{R_\beta}^{ja} + \mathrm{h.c.}\big)\, ,
\ee
where the projectors $P_{u,d}$ project on the up and down components of $ \psi_R^{ja}$ respectively.  

In the most general case  \eq{lag2}  breaks the full group $SU(2)_L\times SU(2)_R\times SU(3)_F$ to the electric charge $U(1)_Q$ in a complete arbitrary way. However, if we think the $h_{\alpha\beta}^{u,d}$ as arising from the vacuum expectation values (VEV) of a field $Y$, they can be written as
$h^{u,d}=\vev{Y^{u,d}}/\Lambda_F $ and, in this case, the Yukawa lagrangian that leads to  \eq{lag2} presents  the extra accidental  global symmetry
$SU(3)^Q_{F_L}\times SU(3)^u_{F_R}\times SU(3)^d_{F_R}$ according to which $Y^{u,d}$ transform as the $(3,\bar{3},1)$ and $(3,1,\bar{3})$ representations, respectively. 
We may now assume that $Y^{u,d}$ develops VEV  only along the  diagonal direction  $SU(3)^u_{F_V}\times SU(3)^d_{F_V}$: in this way,   the Yukawa mass matrices $h^q \; (q=u,d)$ are symmetric and $V_L^q=V_R^q$. Notice that $V_L^{u,d}$ satisfy $V_{CKM}= V_L^{u\dag} V_L^d$, and, thanks to our assumptions,  $V_{CKM}= V_R^{u\dag} V_R^d$.  Thus in the Yukawa sector the  full symmetry $SU(3)^Q_{F_L}\times SU(3)^u_{F_R}\times SU(3)^d_{F_R}\supset SU(3)^u_{F_V}\times SU(3)^d_{F_V}$ is broken  to   $[ U(1)^u]_F^3\times [ U(1)^d]_F^3$. 

It is now simple to identify whether a term of the four-fermion operators   of \eq{4f2} may give rise to flavor violation or not:  those terms that  are invariant under $SU(3)^u_{F_V}\times SU(3)^d_{F_V}$ do not violate flavor because are simultaneously diagonalized with the Yukawa couplings,  the others do.

Indeed, if we classify  quarks  according to their electric charge $q$ and their flavor charge $q_F$ we see that  when in the four-fermion operators flavor indices are contracted between quarks of the same electric charge $q$ automatically the total $q_F$ is zero for each flavor involved.  This happens for exactly 14 operators (each with 2 indices running over the 3 families):
\begin{itemize}
\item 2 operators from $\lambda_1(\lambda_2)$ and 4 operators from $\tilde{\lambda}_1(\tilde{\lambda}_2)$, 2 of which have the same structure of those from $\lambda_1(\lambda_2)$, for a total of  8 operators;
\item 2 operators from $\tilde{\lambda_3}(\tilde{\lambda_4})$ and 3 operators from ${\lambda}_3({\lambda}_4)$, 2 of which have the same structure of those from $\tilde{\lambda_3}(\tilde{\lambda_4})$, for a total of  6 operators.
\end{itemize}
These operators correspond to a subset of possible contractions and give rise to the lagrangian
\bea
\label{opsafe}
{\cal L}_{\psi^4}^{(0)}&= &({\lambda}_{1}+\tilde{\lambda}_{1})\left[ \bar{u}^{a}_{L_A}u^{a}_{R_A} \bar{u}^{b}_{R_B}u^{b}_{L_B} + \bar{d}^{a}_{L_A}d^{a}_{R_A} \bar{d}^{b}_{R_B}d^{b}_{L_B} \right] \nn \\
&+ & \tilde{\lambda}_{1} \left[ \bar{u}^{a}_{L_A}d^{a}_{R_B} \bar{d}^{b}_{R_A}u^{b}_{L_B} + \bar{d}^{a}_{L_A}u^{a}_{R_B} \bar{u}^{b}_{R_A}d^{b}_{L_B} \right] +  {\lambda}_{3}\, \bar{u}^{a}_{L_A}\gamma_\mu u^{a}_{L_A} \bar{d}^{b}_{L_B}\gamma^\mu d^{b}_{L_B} \nn \\
&+&({\lambda}_{3}+\tilde{\lambda}_{3})\left[ \bar{u}^{a}_{L_A}\gamma_\mu u^{a}_{L_A} \bar{u}^{b}_{L_B}\gamma^\mu u^{b}_{L_B} +
\bar{d}^{a}_{L_A}\gamma_\mu d^{a}_{L_A} \bar{d}^{b}_{L_B}\gamma^\mu d^{b}_{L_B}\right] \nn  \\
&+& ({\lambda}_{2}+\tilde{\lambda}_{2})\left[ \bar{u}^{a}_{L_A}u^{b}_{R_A} \bar{u}^{b}_{R_B}u^{a}_{L_B} + \bar{d}^{a}_{L_A}d^{b}_{R_A} \bar{d}^{b}_{R_B}d^{a}_{L_B} \right]  \nn \\
&+ & \tilde{\lambda}_{2} \left[ \bar{u}^{a}_{L_A}d^{b}_{R_B} \bar{d}^{b}_{R_A}u^{a}_{L_B} + \bar{d}^{a}_{L_A}u^{b}_{R_B} \bar{u}^{b}_{R_A}d^{a}_{L_B} \right] +  {\lambda}_{4}\, \bar{u}^{a}_{L_A}\gamma_\mu u^{b}_{L_A} \bar{d}^{b}_{L_B}\gamma^\mu d^{a}_{L_B} \nn \\
&+&({\lambda}_{4}+\tilde{\lambda}_{4})\left[ \bar{u}^{a}_{L_A}\gamma_\mu u^{b}_{L_A} \bar{u}^{b}_{L_B}\gamma^\mu u^{a}_{L_B} +
\bar{d}^{a}_{L_A}\gamma_\mu d^{b}_{L_A} \bar{d}^{b}_{L_B}\gamma^\mu d^{a}_{L_B}\right]  \, , 
\eea
where $A,B$ are mass eigenstates indices, while $a,b$ color indices (notice the different color contractions in the various similar operators).  This lagrangian has, as before in \eq{4f2} 8 coefficients and 14 operators. Each of these operators generates 6 terms once the 3 families are included.

On the contrary when  flavor indices are contracted between quarks that have $|\Delta q|=1$ the operators   present a total $q_F \neq 0$.  This is the case for 6 operators
\begin{itemize}
\item 2 operators from $\lambda_1(\lambda_2)$, for a total of  4 operators;
\item 1 operator from $\tilde{\lambda_3}(\tilde{\lambda_4})$  for a total of 2 operators.
\end{itemize}
The corresponding operators become 4 flavor-index tensors, the structure dictated by the $V_{CKM}$ entries according to
\bea
\label{opnosafe}
{\cal L}_{\psi^4}^{(1)}&=&\lambda_{1}\left[ (V_{CKM})_{AB} (V^\dag_{CKM})_{CD} \,(\bar{u}^a_{L_A} d^{a}_{R_B})(\bar{d}^{b}_{R_C} u^b_{R_D}) +
(V^\dag_{CKM})_{AB} (V_{CKM})_{CD} \,(\bar{d}^a_{L_A} u^{a}_{R_B})(\bar{u}^{b}_{R_C} d^b_{R_D})\right] \nn\\
& +&   \tilde{\lambda}_{3}\, (V_{CKM})_{AD} (V^\dag_{CKM})_{BC}\,( \bar{u}^{a}_{L_A}\gamma_\mu u^{a}_{L_B})( \bar{d}^{b}_{L_C}\gamma^\mu d^{b}_{L_D})
\nn \\
&+&\lambda_{2}\left[ (V_{CKM})_{AB} (V^\dag_{CKM})_{CD} \,(\bar{u}^a_{L_A} d^{b}_{R_B})(\bar{d}^{b}_{R_C} u^a_{R_D}) +
(V^\dag_{CKM})_{AB} (V_{CKM})_{CD} \,(\bar{d}^a_{L_A} u^{b}_{R_B})(\bar{u}^{b}_{R_C} d^a_{R_D})\right] \nn\\
& +&   \tilde{\lambda}_{4}\, (V_{CKM})_{AD} (V^\dag_{CKM})_{BC}\,( \bar{u}^{a}_{L_A}\gamma_\mu u^{b}_{L_B})( \bar{d}^{b}_{L_C}\gamma^\mu d^{a}_{L_D})
\, .  \label{xxx}
\eea
For the  operators in \eq{xxx} flavor breaking is manifest since they are characterized by $q_X=\pm1$ with $X=A,B,C,D$. However these operators may mediate FCNC $\Delta F=2$ processes only at one loop. Thanks to the suppression related to the CKM entries  their effect is of the same order, or even smaller, than the SM contributions. 

To summarize: by imposing an appropriate flavor symmetry,   the large number of  four-fermion operators present in the most general case have been grouped in two classes: 
one that includes operators that conserve flavor, the other those that violate it.  These operators multiply different combination of the 8 independent parameters $\lambda_n$ 
and $\tilde \lambda_n$ in \eq{4f2}.   Flavor violating operators are modulated by the square of CKM entries and are therefore suppressed.  When considering $pp$ collisions, 
the flavor-conserving operators in \eq{opsafe} dominate on those in \eq{opnosafe}---which as a consequence  can be neglected. Moreover, among the operators in  \eq{opsafe}, 
the  largest cross sections are given by those involving only the first family in the initial state  and  the first and, possibly the second family, in the final state. 
Therefore, a complete analysis of this minimal model should take into account at least  the 14  operators  in   the lagrangian ${\cal L}_{\psi^4}^{(0)}$ of \eq{opsafe} 
involving the first family to set bounds on the 8 parameters $\lambda_n$ and $\tilde \lambda_n$.

\section{Event generation and analysis of the skeleton model}
\label{sec:data}

Dijet production in proton-proton collisions $pp\to jj+X$ is the best channel to search for quark contact interactions.  In QCD, the jet production rate peaks at large rapidity $y$, because the scattering is dominated by $t$-channel processes. The rapidity is defined as $y =\frac{1}{2}\ln ( E+p_z)/(E-p_z) $, where $E$ is the energy and $p_z$ the $z$-component of momentum of a given particle. On the other hand, quark contact interactions produce a more isotropic angular distribution leading to enhanced jet production at smaller values of $|y|$. For this reason,  searches for contact interactions at the LHC use quantities computed from these dijet rapidity distributions in the high invariant dijet mass ($m_{jj}$) region.


In  current analyses \cite{atlas,cms}, the contact interactions are parametrized by a single standard operator, the one introduced in \cite{Eichten}---where however a larger set of operators was introduced and the single-operator scenario was only advocated as a simplification. As we argued in the previous section, a standard set of operators can be identified. The simulation of the 14 operators and 8 parameters is rather CPU time consuming and here we only  consider the effect of the presence of more than one operator on dijet production by introducing a skeleton model that admits just two four-fermion operators, which we chose to be the first and the third of \eq{4f}. The size of their couplings  are parametrized by the characteristic energy scales $\Lambda_1$ and  $\Lambda_3$. While a full analysis is certainly necessary, such a radical simplification is already sufficient in showing how the presence of more than one operator gives rise to substantial interference effects which modify the bounds on the characteristic energy scale.

The lagrangian of this skeleton model, written in terms of these parameters, reads:
\be\label{skeleton}
{\cal L}={\cal L}_{QCD}+\frac{2\pi}{\Lambda_3^2} \, \left( \bar \psi_L\gamma_\mu \psi_L \bar \psi_L\gamma^\mu \psi_L +\bar \psi_R\gamma_\mu \psi_R \bar \psi_R\gamma^\mu \psi_R\right)+\frac{2\pi}{\Lambda_1^2}\left( \bar \psi_L \psi_R \bar \psi_R \psi_L\right)\,.
\ee 
where the isospin and the color contractions, that can be read out directly from \eq{4f}, are omitted.

This choice of considering just two operators is useful because it reduces the CPU time for the simulation, simplifies the analysis, and shows how it is essential to consider more than one operator.

We use {\tt MADGRAPH V.4.5.0} to simulate LHC dijet production in $pp$ collisions at $\sqrt{s} = 7$ TeV,
Monte Carlo samples are generated for pure QCD and for QCD modified by the new four fermion interaction terms in \eq{skeleton}. Since {\tt MADGRAPH} is a leading-order generator 
and does not support non-renormalizable interactions, we have implemented them effectively, introducing a set of  fictitious gauge interactions acting only on the first quark 
family. In this case, the identification is $\psi=(u\,\,d)$, and the mass of the fictitious gauge boson has been choosen to be very high  ($\sim 100$ TeV).
The generated events are then passed through {\tt PGS}, the detector simulator in which the parameters are set to reproduce the ATLAS detector performance.

In order to restrict the simulation in the kinematical region of interest, we have applied the following cuts at the generator level:
\begin{itemize}
\item $m_{jj}>1000$ GeV\,;
\item $p_T^{j_1},p_T^{j_2}>30$ GeV\,; 
\item $|\eta^{jet}|<2.8$  \,.
\end{itemize}
The pseudorapidity is defined as $\eta = -\ln(\tan \theta/2)$, where $\theta$ is the angle between the jet and the beam direction in the laboratory frame. 

We have generated  Monte Carlo samples for different values of the energy scales $\Lambda_1$ and $\Lambda_3$ between 1 and 10 TeV. 

Given the cuts described above, the Monte Carlo leading order cross section for each choice of the parameters turns out to be $\sigma \simeq 5\times 10^3$ pb.
An integrated luminosity of 1 fb$^{-1}$ has been generated for each of the points in the ($\Lambda_1$, $\Lambda_3$) plane described above.


The variable $\chi$ is the quantity used for the angular distribution study. It is defined as a function of the rapidities of the two highest $p_T$ jets in the event, $y_1$ and $y_2$:
\be
\label{chi_def}
\chi=\exp(2|y^*|)\,,
\ee
In the massless particle limit the center-of-mass (CM) rapidity $y^*=(y_1-y_2)/2$ is used to determine the partonic CM angle $\theta^*$ given the relation 
$y^*=\frac{1}{2}\rm{ln}\left( \frac{1+|\rm{cos}\,\theta^*|}{1-|\rm{cos}\,\theta^*|}\right)$. Another variable used in the following, derived from the rapidities of the two jets, 
is $|y_B|=(y_1+y_2)/2$. The $\chi$ variable is  useful for the comparison of angular distribution predicted by new processes with those of QCD. 
In QCD, gluon exchange diagrams have approximately the same angular dependence as Rutherford scattering and so the $dN/d\chi$ distribution is constant in $\chi$. 
On the other hand, the angular distributions of new processes are more isotropic, leading to additional events in the low $\chi$ region. 
Subleading diagrams in QCD can also rise slightly the distribution at low $\chi$. 

The second important kinematic variable  is the dijet invariant mass $m_{jj}$, which is also the CM energy of the partonic system. It is computed from the two jet 
four-vectors as 
\be
m_{jj}=\sqrt{(E^{j1}+E^{j2})^2-(\mathbf{p}^{j1}+\mathbf{p}^{j2})^2} \, ,
\ee
where $E$ and $\mathbf{p}$ are the energy and momentum of the jets.

Events with at least two jets are retained if the highest $p_T$ jet satisfies $p_T^{j1}> 60$ GeV and the second highest one satisfies  $p_T^{j2}> 30$ GeV. This asymmetric 
thresholds avoid suppression of events where a third jet has been radiated, while $30$ GeV threshold ensures that reconstruction is fully efficient for both leading jets. 
Events with an additional poorly measured jet with $p_T > 15$ GeV are vetoed to avoid possible incorrect identification of the two leading jets. 
In addition, $\chi$ distribution are accumulated only for events for which $|y_B|<1.10$ and  $|y^*|<1.70$.

\begin{figure}[ht!]
\begin{center}
\includegraphics[width=4in]{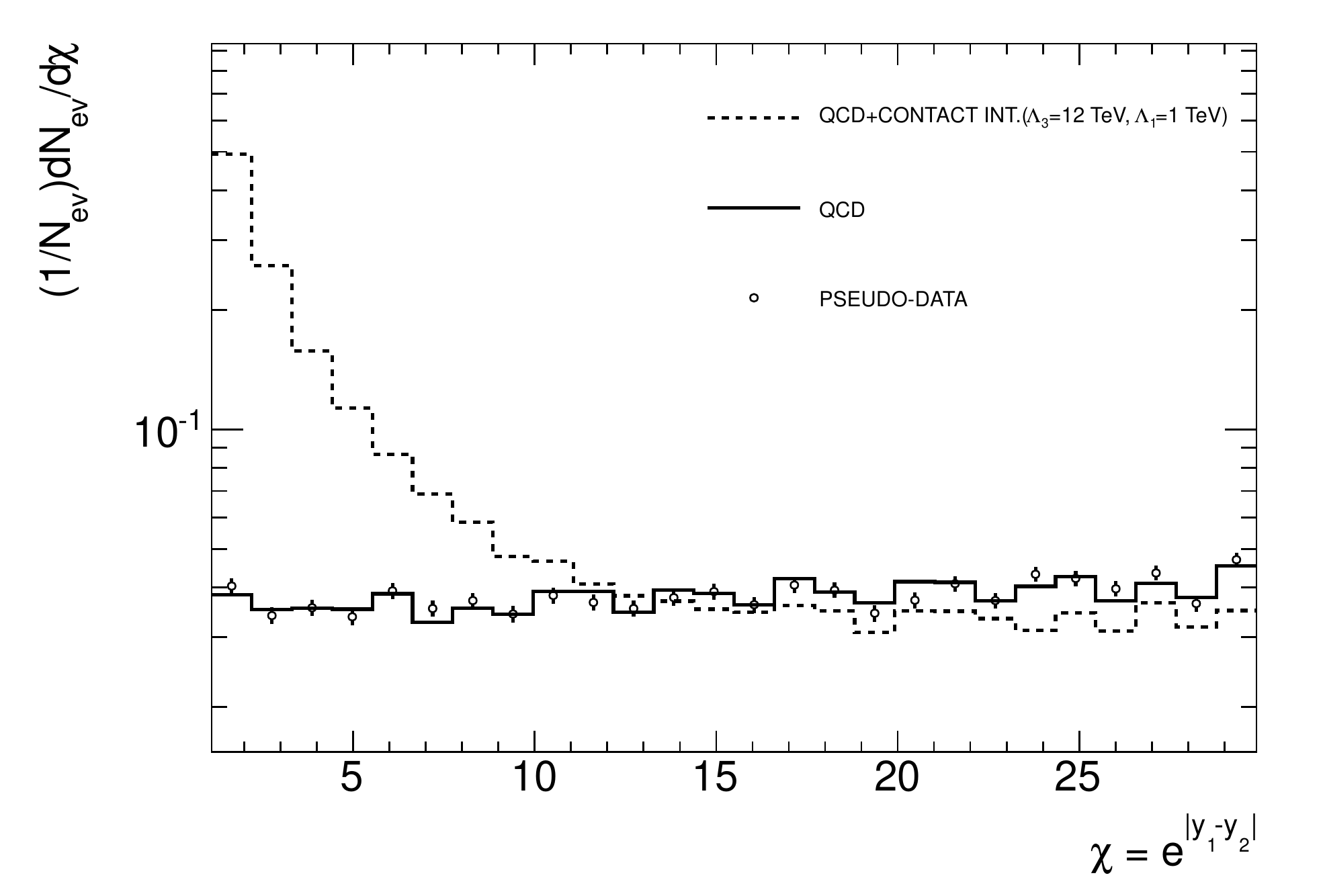} 
\caption{\small $\chi$ distribution, as defined in \eq{chi_def} for events with $m_{jj} \ge 1200$ GeV.  Empty points represent the $\chi$ distribution for pseudo-data, 
the solid line for pure QCD while the dotted line for one specific point ($\Lambda_{1} = 1$ TeV, $\Lambda_{3} = 12$ TeV) in the generated grid. 
The various distributions are normalized to an integrated luminosity of 200 pb$^{-1}$.}
\label{chi_dist}
\end{center}
\end{figure}

The measure of the isotropy in the dijet distribution, introduced in \cite{atlas}, is given by the variable $F_\chi$. It measures the fraction of dijets produced centrally 
versus the total number of observed dijets in a specified dijet mass range:
\be
F_\chi=\frac{N_{\rm{events}}(|y^*|<0.6)}{N_{\rm{events}}(|y^*|<1.7)}\, ,
\label{fchi_def}
\ee
where $N_{\rm{events}}$ is the number of candidate events within the $y^*$ interval. The central region which is expected to be most sensitive to new physics is defined by the interval $|y^*|<0.6$ and corresponds to $\chi < 3.32$, while $|y^*|<1.7$ extends the angular range to $\chi < 30$.


\begin{figure}[ht!]
\begin{center}
\includegraphics[width=4in]{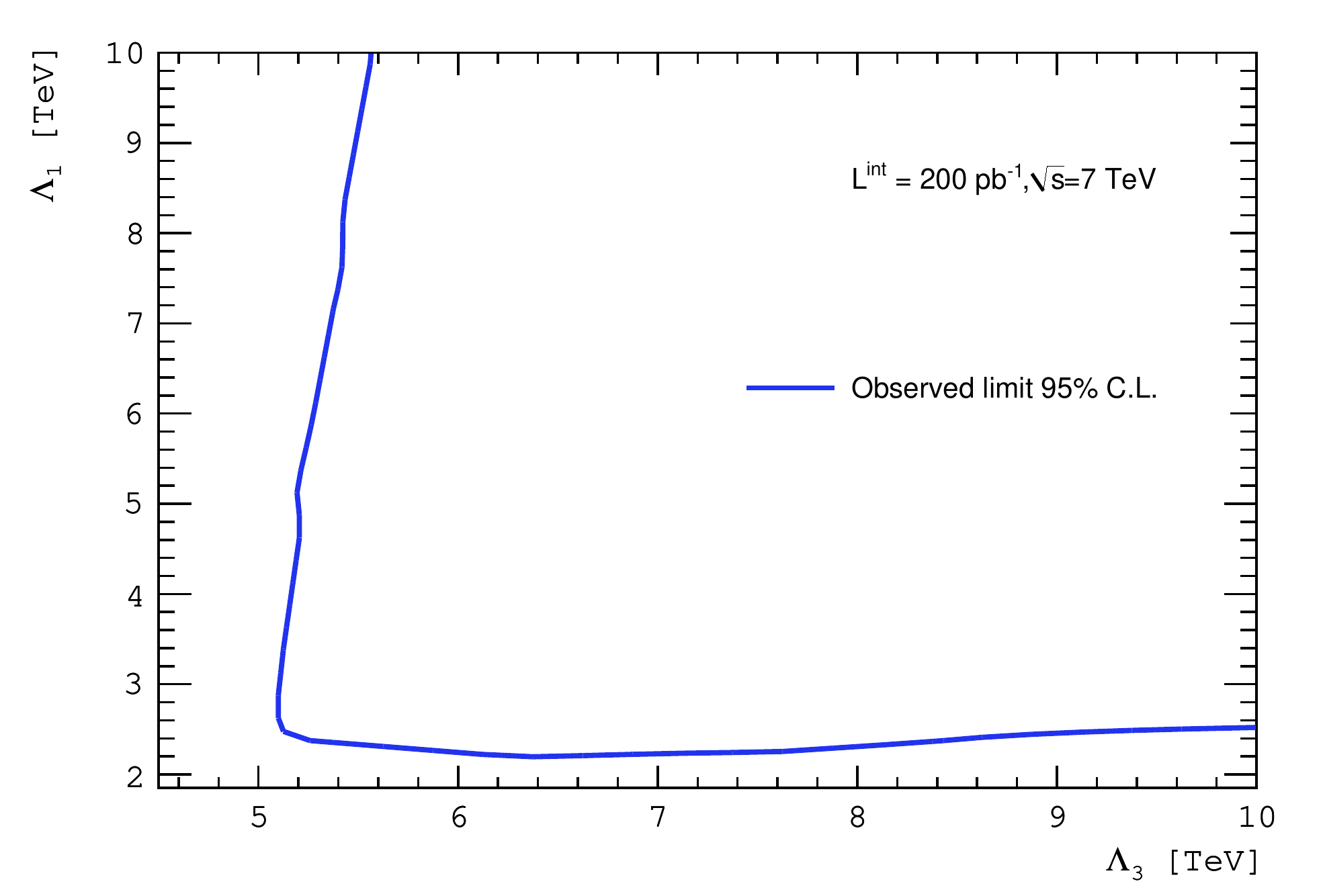} 
\caption{\small Isotropy of the dijet distribution as given by the variable $F_\chi$. The  area outside the curve represents values above  the value computed from the QCD pseudo-data. 
The area inside represents---at the 95\% confidence level---values compatible with the pseudo-data obtained for different energy scales  $\Lambda_1$ and  $\Lambda_3$ of the 
contact interactions.  The lower bounds for  $\Lambda_1$ and  $\Lambda_3$ are, respectively, 2.2 and 5.1 TeV.
\label{effe_chi}}
\end{center}
\end{figure}

The presence of possible contact terms  is tested for each value of $\Lambda_1$
and $\Lambda_3$ in the highest dijet mass bin: $m_{jj}\ge 1200$ GeV. For a given  pair of values of the energy scales, the corresponding value of $F_\chi$  is obtained starting
from the $\chi$ distribution as in Fig.~\ref{chi_dist}. We have generated a QCD Monte Carlo sample, corresponding to an integrated luminosity of 200 pb$^{-1}$, to be used as a
pseudo-data sample. In Fig.~\ref{chi_dist}, the $\chi$ distribution is shown for dijet events passing the selection described above with the additional
constraint that the invariant mass of the two hardest jets is larger than 1200 GeV. The pseudo-data sample (empty points) is well described by the QCD (solid line), while the
contribution of a contact interaction term (dotted line), corresponding to the point ($\Lambda_{1} = 1$ TeV, $\Lambda_{3} = 12$ TeV) in the grid, clearly shows a peak in the low
$\chi$ region, giving then a larger values for $F_\chi$ (defined in \eq{fchi_def}) with respect to both pseudo-data and QCD samples.
A full set of pseudo-experiments, has then been made for each of the points in the grid in order to 
construct one-sided 95\% confidence level. 

The result of this analysis is shown by the contour plot of Fig.~\ref{effe_chi}. 
The value of $F_\chi$ extracted from the pseudo-data  represents in Fig.~\ref{effe_chi} the level below which the contact interactions are compatible with pseudo-data. 
The values of $\Lambda_1$ and $\Lambda_3$ which satisfy---at the 95\% confidence level---this bound are represented by the area inside the curve.
By inspection, we find that the  lower bounds on the contact interaction scale are given by the values $\Lambda_1= 2.2$ TeV and $\Lambda_3= 5.1$ TeV, respectively. 

The standard one-operator analysis---which corresponds to taking $\Lambda_1$  very large---would give a limit $\Lambda_3=\Lambda \approx 5.6$ TeV which corresponds to the upper margin of 
the not excluded area in the contour plot in Fig.~\ref{effe_chi}. The bounds we find in the case of two operators are weaker than this one because of  interference effects.

If one insists in having a unique  energy scale $\Lambda$ even in the presence of more operators, it possible to provide it by combining the characteristic scales $\Lambda_n$ by means of, for instance,  the definition
\be
\frac{1}{\Lambda^2} = \sum_{n=1}^{8} \frac{1}{\Lambda_n^2}  \, . \label{ll}
\ee
This definition has the advantage of providing a bound close to the lowest one and to go into the single-operator limit when all scales but one are taken to be large.
In our skeleton model with just two operators  the definition in \eq{ll} gives a bound $\Lambda= 2.0$ TeV---a value again weaker than what found in the single-operator analysis.

\acknowledgments

We thank O. Mattelaer and C. Degrande for  help with {\tt MADGRAPH}; G. Choudalakis, F. Ruehr and F.Meloni for their helpful suggestions in performing the analysis. 


\end{document}